\documentclass[CEJP,PDF]{cej} % use PDF command to enable PDFLaTeX driver
\usepackage{layout}
\usepackage{amsmath}
\usepackage{textcomp}

\title{Nonextensive thermal sources of cosmic rays}

\articletype{Research Article}

\author{Grzegorz Wilk\inst{1}\email{wilk@fuw.edu.pl},
        Zbigniew W\l odarczyk\inst{2}\email{wlod@pu.kielce.pl}}

\institute{\inst{1} The Andrzej So\l tan Institute for Nuclear
                   Studies, Theoretical Physics Department,\\
                   ul. Ho\.za 69, 00-681 Warsaw, Poland,
           \inst{2} Institute of Physics, Jan Kochanowski
                    University,\\
                    \'Swi\c{e}tokrzyska 15, 25-405 Kielce, Poland,
            }

\abstract{The energy spectrum of cosmic rays (CR) exhibits
power-like behavior with a very characteristic "knee" structure.
We consider a possibility that such a spectrum could be generated
by some specific nonstatistical temperature fluctuations in the
source of CR with the "knee" structure reflecting an abrupt change
of the pattern of such fluctuations. This would result in a
generalized nonextensive statistical model for the production of
CR. The possible physical mechanisms leading to these effects are
discussed together with the resulting chemical composition of the
CR, which follows the experimentally observed abundance of nuclei.
}


\begin{document}
\maketitle

%% ###################################################################

\section{\label{sec:I}Introduction}

The energy spectrum of cosmic rays (CR) has characteristic
power-like behavior with a "knee" structure (plus some other less
prominent features) and remains constantly matter of hot debate
(see \cite{CR-origin} and references therein). It could reflect
the action of different regimes of diffusive propagation of CR in
the Galaxy combined with its different chemical composition, but
it could also be due to some, so far unspecified, property of the
production processes within the source of the CR itself. In this
work we shall consider this possibility assuming that CR are
produced following a generalized nonextensive thermal approach
\cite{Tsallis,WWTq}. Actually, nonextensive statistical mechanics
\cite{Tsallis} has been applied to CR before: in \cite{previous1}
the "knee" structure was attributed to the crossover between the
assumed fractal-like thermal regimes of CR propagation
(characterized by different temperatures $T$ and nonextensive
parameters $q$), whereas in \cite{previous2} the possible
nonextensive thermal features of CR flux have been investigated
but only up to the "knee" region, the origin of which was not
discussed. In both cases the obtained values of temperatures were
much too high to be accommodated by any known physical mechanism.
In this paper we propose a mechanism which is apparently capable
to describe the whole spectrum of CR, including the "knee" region,
using physically reasonable values of temperature of the source of
CR. The observed power-like behavior of the energy spectrum of CR
is attributed (as in \cite{previous2}) to fluctuations of the
temperature in the source producing CR and the occurrence of
"knee" (cf., Fig. \ref{fig1}a) is connected with some abrupt
change of this fluctuation pattern \cite{WW}, visualized by a
dramatic change in the nonextensivity parameter observed Fig.
\ref{fig1}b. However, to keep the temperature of the CR source
acceptably low, one has to allow additionally for energy transfer
to the production region; this is assumed to proceed through the
mechanism proposed in \cite{WWTq} and is characterized by some
effective temperature $T_{eff}$ \footnote{Actually, this mechanism
was originally invented to describe some features of heavy ion
collisions \cite{hi} in which energy was transferred out of the
system; in the case of the CR it is transferred towards the
system.}. These points summarize what we call a generalized
nonextensive thermal approach (GNTA), which we shall now describe
in more detail.

It must be stressed at this point that, for the sake of clarity of
presentation, we consider in what follows only a very simplified
situation. Namely, we assume that GNTA is, for a moment, the only
mechanism of production of the CR present. It must be realized
that in reality GNTA  would have to be incorporated into many
other possibilities considered in the usual analysis of CR (and
listed, for example, in \cite{CR-origin}).

The organization of our paper is as follows. In the next Section
we present some basic considerations concerning nonextensive
statistics and CR, out of which the discussion of the chemical
composition of CR seen from that point of view is a new element
here. Section \ref{sec:III} contains our results and their
physical interpretation in terms of some specific properties in
the superfluid stages of neutron stars supplied by the proposition
of introducing phenomenologically the energy transfer to CR
(described by some effective temperature $T_{eff}$ and needed to
assure the consistency of obtained parameters). In Section
\ref{sec:IV} we discuss the influence of acceleration and
propagation of CR on energy spectra and composition. A summary and
concluding remarks are presented in Section \ref{sec:V}.

\section{\label{sec:II}Basic elements of nonextensive statistics and Cosmic Rays}

\subsection{\label{sec:IIA}Generalities}

Nonextensive statistical mechanics as proposed and developed in
\cite{Tsallis} is based on the generalized entropy functional
(Tsallis entropy),
\begin{equation}
S_q \, =\, - \frac{\int dE\, P^q(E)\, -\, 1}{q-1}\, \, .
\label{eq:Sq}
\end{equation}
Its maximization under appropriate constrains yields a
characteristic power-like distribution ($q$-exponential
distribution, $\exp_q(\dots)$):
\begin{equation}
P_q(E) = \frac{2-q}{T} \exp_q\left( - \frac{E}{T}\right) =
\frac{2-q}{T}\, \left[ 1 - (1-q) \frac{E}{T}
\right]^{\frac{1}{1-q}}. \label{eq:P_q}
\end{equation}
For $q \rightarrow 1 $ one recovers the usual
Boltzmann-Gibbs-Shannon (BGS) entropy and the usual exponential
distribution. This equilibrium distribution can alternatively be
obtained by solving the following differential equation,
\begin{equation}
\frac{dP(E)}{dE} = - \frac{P^q(E)}{T}. \label{eq:difeqT}
\end{equation}
The extended version of this equation with two terms
(accommodating two different values of $q$ and $T$) has been used
with apparent success in \cite{previous1} to describe the flux of
CR. The "knee" appears there as a crossover between two
fractal-like thermal regimes characterized by $(T,q)$ and
$(T',q')$. However, the values of temperatures obtained there
($(T,T') \sim 100 \div 1000$ MeV) are uncomfortably high and
cannot be attributed to any known mechanism of CR production.

On the other hand, there is growing evidence that a nonextensive
formalism applies most often to nonequilibrium systems with a
stationary state that possesses strong fluctuations of the inverse
temperature parameter $\beta = 1/T$ \cite{WW,Biro} (cf., also
\cite{Add}). In fact, fluctuating $\beta$ according to gamma
distribution (cf. Eq. (\ref{eq:Gamma}) below) with variance
$Var(\beta)$ results in a power like distribution (\ref{eq:P_q})
with the deviation of the nonextensivity parameter $q$ from unity
being given by the strength of these fluctuations,
\begin{equation}
q-1 \, =\, \frac{Var(\beta)}{\langle \beta\rangle^2}.
\label{eq:varbeta}
\end{equation}
This observation was used in \cite{previous2} to describe the
energy spectrum (but only up to the "knee" region). Again,
although the results were reasonably good the estimated
temperature $T\sim 170$ MeV is far too high. This was because the
author insisted on the description of the whole range of energy
spectrum up to the "knee" region, including its very low energy
part, which is, however, governed mainly by the geomagnetic
cut-off and diffusion effects and should therefore be considered
separately (it is thus not covered by our approach).

\subsection{\label{sec:IIB}Energy spectrum}

\begin{figure}[t]
\begin{center}
\includegraphics [width=9cm]{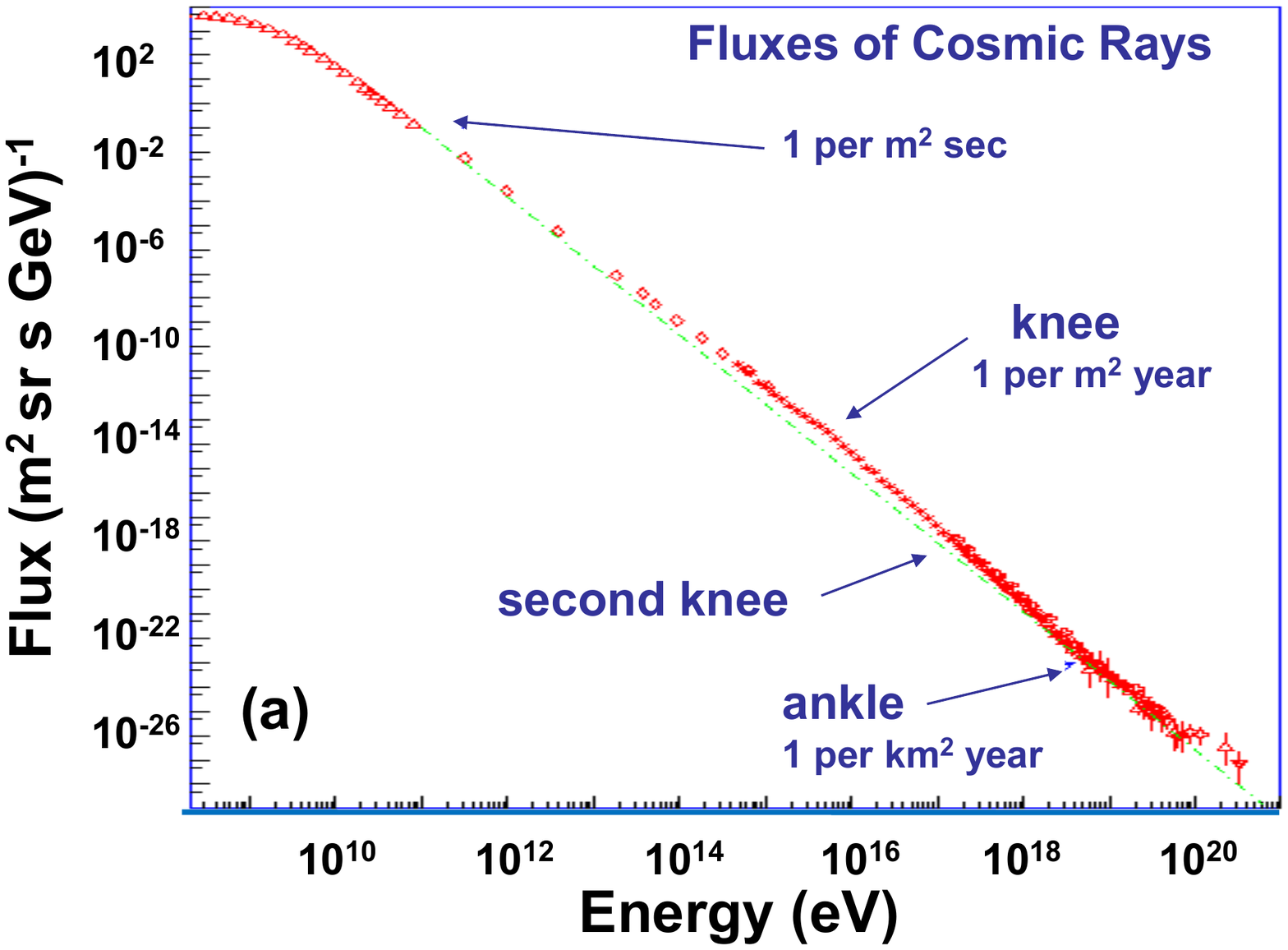}\hspace{-2.5cm}
\includegraphics [width=9cm]{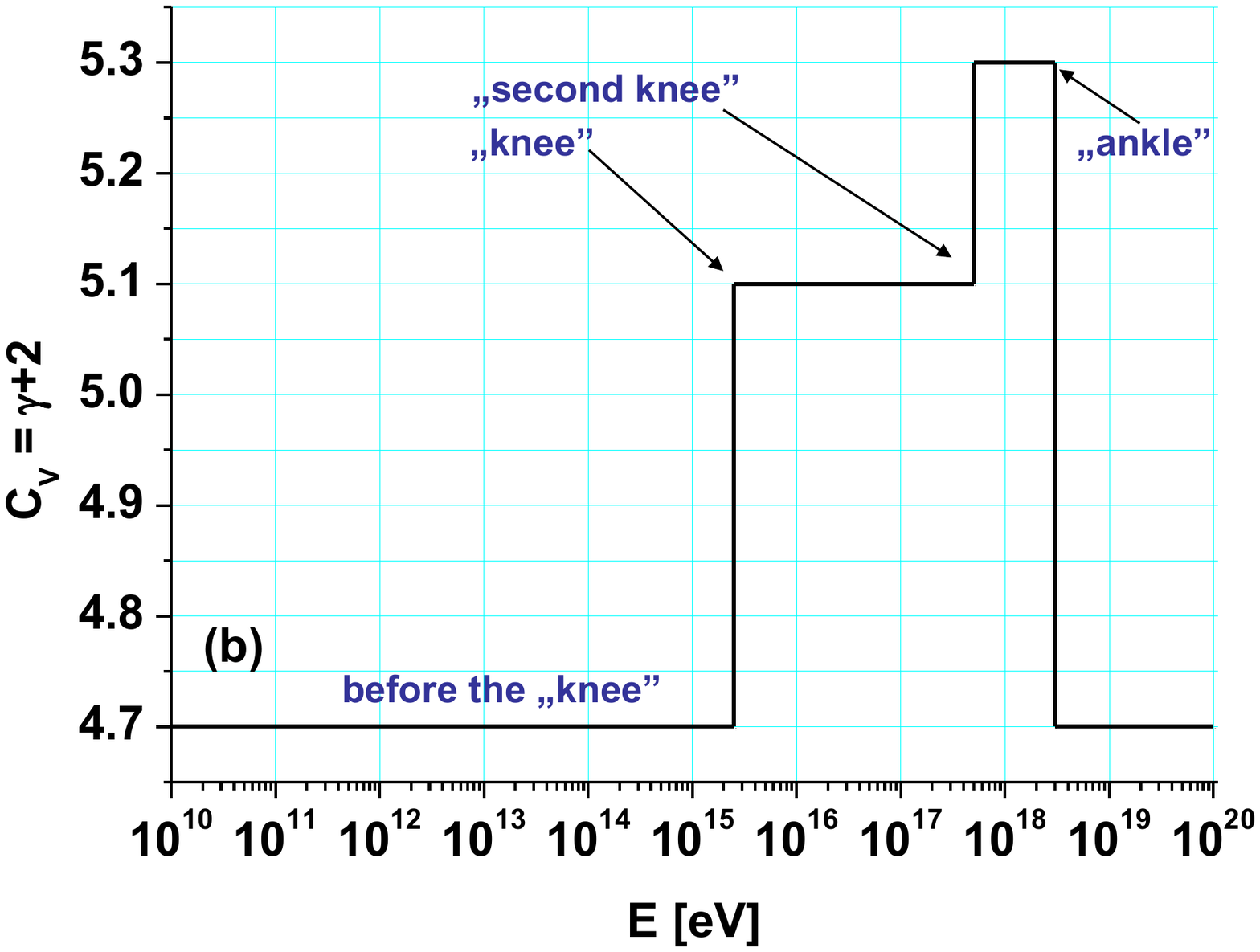}
\vspace{-1.5cm} \caption{(Color online) $(a)$ Schematic view of
the observed CR energy spectrum \cite{CR-origin} with
characteristic features clearly indicated. $(b)$ The pattern of
fluctuations obtained from the above by means of nonextensive
statistics approach and represented by the heat capacity $C_V$ as
given by Eq. (\ref{eq:C_V}) (see text for details). Notice that
previous irregularities are now dramatically enhanced.}
\label{fig1}
\end{center}
\end{figure}

Treating CR as relativistic particles (for which rest mass $m$ can
be neglected) their energy is $E\sim p$ and the density of states
is that of an ideal gas in three dimensions, $\Omega (E) \propto
E^2$. The corresponding energy spectrum $\Phi(E)$ is then
\begin{equation}
\Phi(E)\, =\, N_0\, E^2\, P(E), \label{eq:flux}
\end{equation}
where  $N_0$ is normalization factor. For $P(E)$ given by Eq.
(\ref{eq:P_q}) we have, for $E >> T$, power spectrum
\begin{equation}
\Phi(E) \propto E^{-\gamma}; \qquad \gamma = \frac{3-2q}{q-1}.
\label{eq:p_spectrum}
\end{equation}
As seen in Fig. \ref{fig1}a it changes in the region named "knee"
where  the slope $\gamma_1 \simeq 2.7$ at energies below $\sim
10^{15}$ eV and $\gamma_2 \simeq 3.1$ above it. In the language of
the nonextensivity parameters it would mean that $q_1 = 1.213$
before and $q_2 = 1.196$ after the "knee". For $q$ understood as a
measure of fluctuations, as it is the case in our paper, one
therefore witnesses at the "knee" a change of fluctuation pattern.

\subsection{\label{sec:IIC}The chemical composition}

\begin{figure}[t]
\begin{center}
\includegraphics [width=10cm]{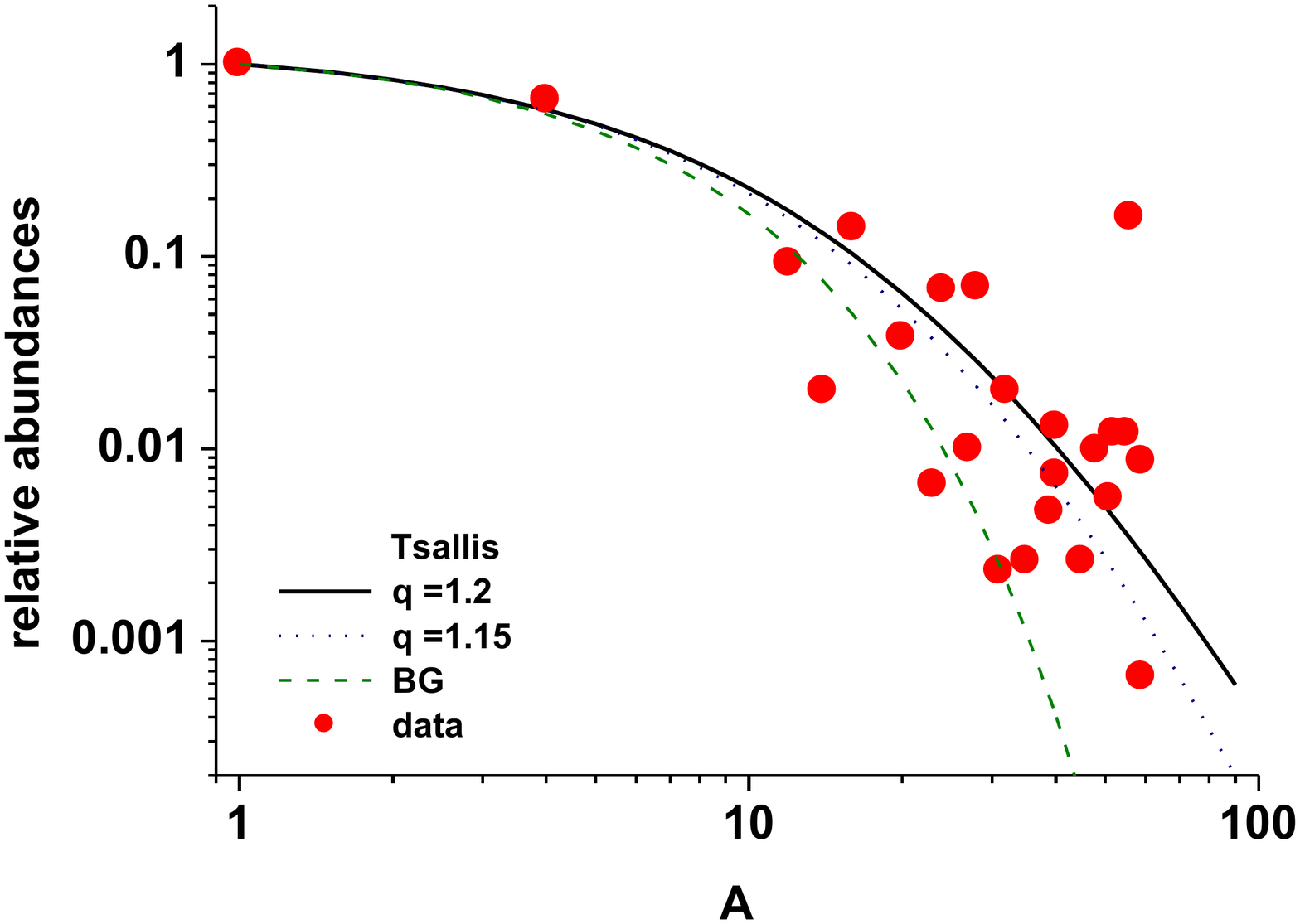} \vspace{-1.5cm}
 \caption{(Color online) The chemical composition of CR (relative to hydrogen
          at $1$ TeV) corresponding to nonenextensive picture
          advocated in this work.  Data points are from
          \cite{Abundances}. }
 \label{fig2}
 \end{center}
\end{figure}

If the energy distribution of CR follows the Tsallis formula
(\ref{eq:P_q}), it is natural to expect that also the abundance of
nuclei with mass $A$ will follow the same pattern. We therefore
expect that
\begin{equation}
\omega (A) \propto \left[ 1 - (1 -
q)\frac{A\varepsilon}{T}\right]^{\frac{1}{1-q}},
\label{eq:abundance}
\end{equation}
where $\varepsilon$ is the average energy of nucleon equal to
$\varepsilon = \frac{3}{5}E_F$. The typical value of the Fermi
energy for nucleus consisting of $A$ nucleons (distributed in
sphere of radius $R = 1.25 A^{1/3}$) is $E_F \simeq 30$ MeV, it
means then that $\varepsilon \simeq 22$ MeV. In Fig. \ref{fig2} we
show $\omega(A)/\omega(A=1)$ for $\varepsilon = 22$ MeV and $T =
100$ MeV as function of $A$ for two values of $q$: $q = 1.2$ and
$q = 1.15$. As one can see the sensitivity to $q$ is rather weak
\footnote{In fact, the relative abundance shown in Fig. \ref{fig2}
depends on the temperature chosen. However, reasonable fits are
possible only in the limited temperature interval, $60$ MeV $< T <
160$ MeV, and for the nonextensivity parameter satisfying roughly
the relation $q = 1.5 - 0.29 T/100$. Our fit is for the same
parameters $(T,q)$ as used to describe the energy spectrum .}. The
average mass number (for the spectrum $\omega(A)$) is
\begin{equation}
\langle A\rangle = \frac{2 - q + T/\varepsilon}{3 - 2q}.
\label{eq:Comp}
\end{equation}
Numerically evaluated $\langle \ln(A)\rangle$ equals $1.82$ below
the "knee" (for greater $q$) and $1.78$ above the "knee" (for
smaller $q$) and shows that predicted changes of chemical
composition (due to changes of spectral index) in the "knee"
region are negligible. Notice that the upper limit estimation for
the usual BG distribution (corresponding to $q = 1$ here) leaves
the majority of points for large values of $A$ well above the
curve. From this point of view our prediction is much better
(although, the thermal model alone is already able to provide
quite reasonable chemical composition of CR) \footnote{A remark of
caution is in order here. Our $\langle A\rangle$ agrees with
observations in the low energy region but the observed composition
shows changes with energy. On the other hand, we show small
changes connected with the change of the spectral index $\gamma$
and this could be caused by some other mechanism influencing
chemical composition which we have not accounted for. The most
important is problem of energy dependence of the effective
temperature $T_{eff}$ (cf., Eq. (\ref{eq:Teff})). In fact, we do
not know $T_0$ and cannot say (numerically) how $T_{visc}$ changes
with the energy. Instead, we just put roughly $T_{eff} = 100$ MeV.
A more exact analysis of experimental data, including energy
dependence of the chemical composition, would be helpful in
estimation of $T_0$ itself. We plan to address this problem
elsewhere.}.

One should keep in mind that the relative dissemination of
nuclides shows many intriguing features which should be connected
with specific properties of nuclei, cf., for example,
\cite{properties}. We close this section by stressing that our
$\omega(A)$ as given by Eq. (\ref{eq:abundance}) (and therefore
represented by a continuous curve) does not account for any
differences between nuclei. In fact, it even does not describe the
abundance in the source. It describes only the emissive power of
thermal source and shows that this factor (neglecting differences
between particular nuclei) determines the global characteristics
of the observed spread of nuclides.

\section{\label{sec:III}Results}

\subsection{\label{sec:IIIA}Temperature fluctuations}

As mentioned above, the special role in converting an exponential
distribution to its $q$-exponential counterpart play fluctuations
of the inverse temperature $\beta$ described by a gamma function
\cite{WW,Add},
\begin{equation}
f(\beta) = \frac{\mu}{\Gamma(\nu)}\left( \mu \beta\right)^{\nu
-1}\, \exp\left( - \beta \mu\right) \label{eq:Gamma}
\end{equation}
where $\mu^{-1} = \beta_0(q-1)$ and $\nu^{-1} = q-1$. There are a
priori at least two scenarios leading to such $f(\beta)$: $(i)$
one can have many sources with different temperatures, the number
of which is distributed that way or $(ii)$ one can have
temperature fluctuations in small parts of a source. The first
possibility is, however, rather unlikely because in this case
either one would have to accept sources with unphysically large
and small temperatures or else use the temperature distribution in
some limited domain, i.e., work with a truncated version of gamma
distribution. This, however, would result in a very characteristic
rapid break in the energy spectrum. This is not observed in the
experiment.

We shall therefore concentrate on the second possibility. To
illustrate it, suppose one has a thermodynamic system, different
small parts of which have locally different temperatures, i.e.,
its temperature understood in the usual way fluctuates. Let
$\xi(t)$ describes the stochastic changes of temperature in time
and let it be defined by the white Gaussian noise ($\langle
\xi(t)\rangle = 0$ and $\langle \xi (t) \xi (t+\Delta t)\rangle  =
2 D \delta(\Delta t)$). The inevitable exchange of heat which
takes place between the selected regions of our system leads
ultimately to an equilibration of temperature and, as shown in
\cite{WW}, the corresponding process of heat conductance
eventually leads to the gamma distribution (\ref{eq:Gamma})
mentioned before with variance (\ref{eq:varbeta}) related to the
heat capacity $C_V$ of this system (expressed in units of
Boltzmann constant $k_B$, which we put equal unity in what
follows) by \cite{S}
\begin{equation}
\frac{Var(\beta)}{< \beta>^2} = \frac{1}{C_V} \label{eq:S}
\end{equation}
and we have finally that
\begin{equation}
C_V = \frac{1}{q-1} = \gamma + 2, \label{eq:C_V}
\end{equation}
where we have used Eq. (\ref{eq:p_spectrum}) connecting the
spectral index $\gamma$ of the energy spectrum with the
nonextensivity parameter $q$. In this way, we come directly to the
possible physical interpretation of the nonextensivity parameter
which allows us to translate the pattern observed in Fig.
\ref{fig1}a into that shown in Fig. \ref{fig1}b. Here $C_V$, as
given by Eq. (\ref{eq:C_V}), is shown as a function of energy at
which we observe the essential influence of given $C_V$ on the
slope of energy spectrum. As will be discussed in detail below in
Sec. \ref{sec:IIIB}, although $C_V$ as such is energy independent
it can change with temperature $T$ in the source and changes
abruptly at some temperature $T_c$ (identified, for example, with
$T_{cut}$ in Sec. \ref{sec:IIIB}). These changes of $C_V$ result
in changes of the slope of the observed energy spectrum at $E \sim
2T_{c}$).

To summarize this part: In what follows, we shall concentrate
mainly on the change of fluctuation pattern in the "knee" region
and we shall argue that it could indicate an abrupt change in the
heat capacity in the source of CR of the order of $C_2/C_1 =
1.09$. Notice that this change is much more pronounced and
dramatic than the corresponding change of slope in the "knee"
region observed in Fig. \ref{fig1}a.

\subsection{\label{sec:IIIB}Possible physical interpretations of
the fluctuation pattern}

Can one expect something of this kind to happen in the
astrophysical environment of the CR? In what follows we shall
argue that, indeed, one can. Let us first notice that the subject
of temperature fluctuations in astrophysics is a much-discussed
problem nowadays. Its  effect on the temperatures empirically
derived from spectroscopic observations was first investigated in
\cite{P} whereas in \cite{13a,13b,13c,13d} it was shown that
temperature fluctuations in photoionized nebulea have great
importance to all abundance determinations in such objects. This
means that discussion of the heat capacity or, equivalently, the
behavior of the parameter $q$ defining the energy spectrum, is
fully justified. In what follows we shall concentrate on the
problem of the heat capacity of astrophysical objects, in
particular in neutron stars, concentrating on some peculiarities
connected with their description in terms of Fermi liquids.

We start with fluid/superfluid transitions in such systems and
their effect on the heat capacity.  In neutron stars one observes
the following feature. The total specific heat of their crust,
$C$, is the sum of contributions from the relativistic degenerate
electrons, from the ions and from degenerate neutrons. In the
temperature that can be reached in the crust of an accreting
neutron star (which is of the order of $T \sim5\cdot 10^8$ K and
is below the Debaye temperature $T_D \sim 5\cdot 10^9$ K) we have
$C_{ion} < C_e < C_n$. When the temperature drops below the
critical value $T = T_C$ the neutrons become superfluid and their
heat capacity $C^{sf}_n$ increases \cite{HC1,HC2},
\begin{eqnarray}
\frac{C^{sf}_n}{C_n} \simeq 3.15\frac{T_C}{T} \exp\left( -
1.76\frac{T_C}{T}\right)\cdot \left[2.5 - 1.66
\left(\frac{T}{T_C}\right) + 3.68 \left(\frac{T}{T_C}\right)^2
\right] .\label{eq:C/C}
\end{eqnarray}
At  $T \sim 0.7 T_C$ we have $C^{sf}_n \sim 1.1 C_n$ what
corresponds to the changes of spectral index by $\Delta \gamma
\sim 0.5$. To summarize: one witnesses here an abrupt change in
the heat capacity at some temperature, i.e., a phenomenon we were
looking for.

\begin{figure}[t]
\begin{center}
\includegraphics[width=9cm]{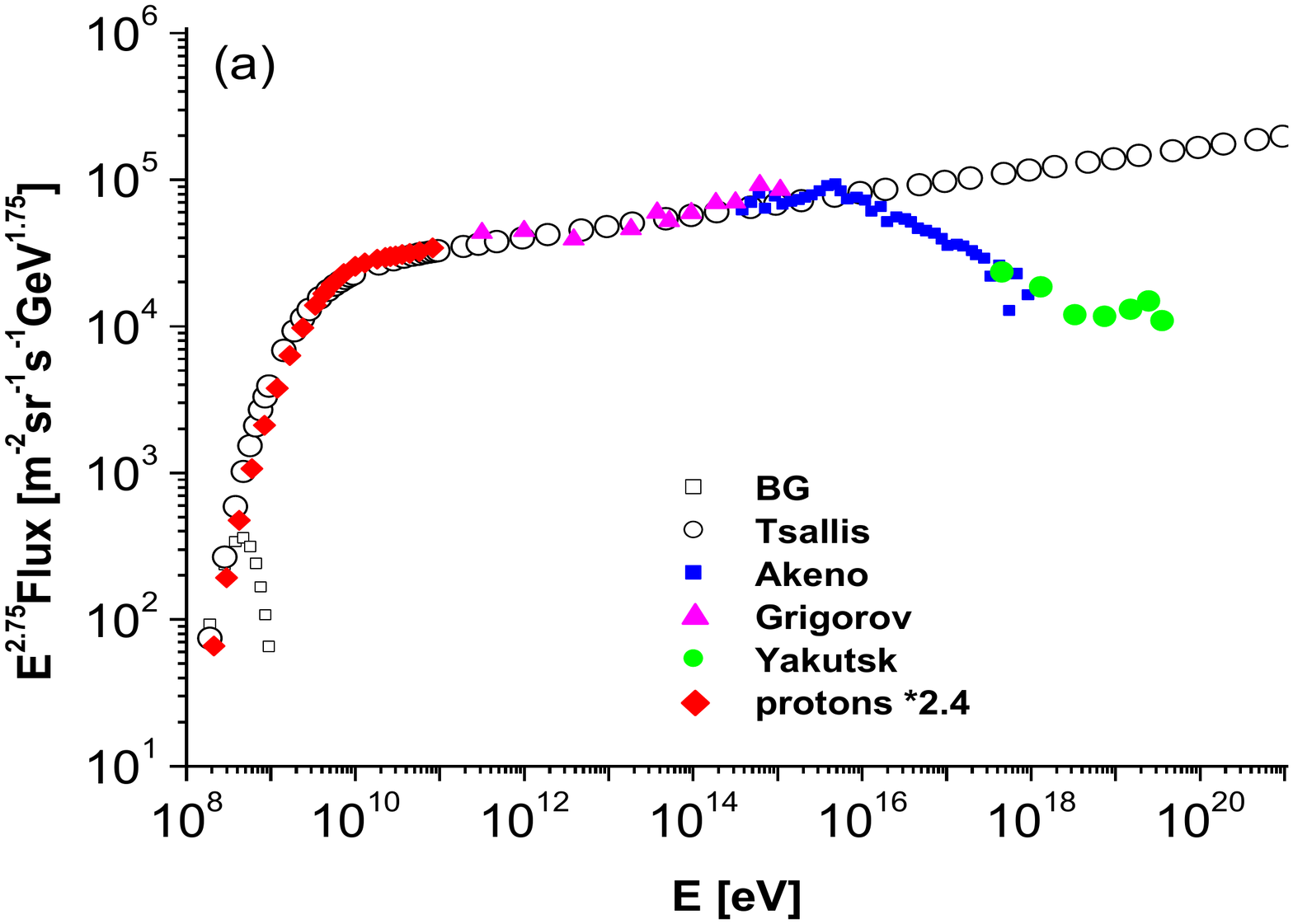}\hspace{-2.5cm}
\includegraphics [width=9cm]{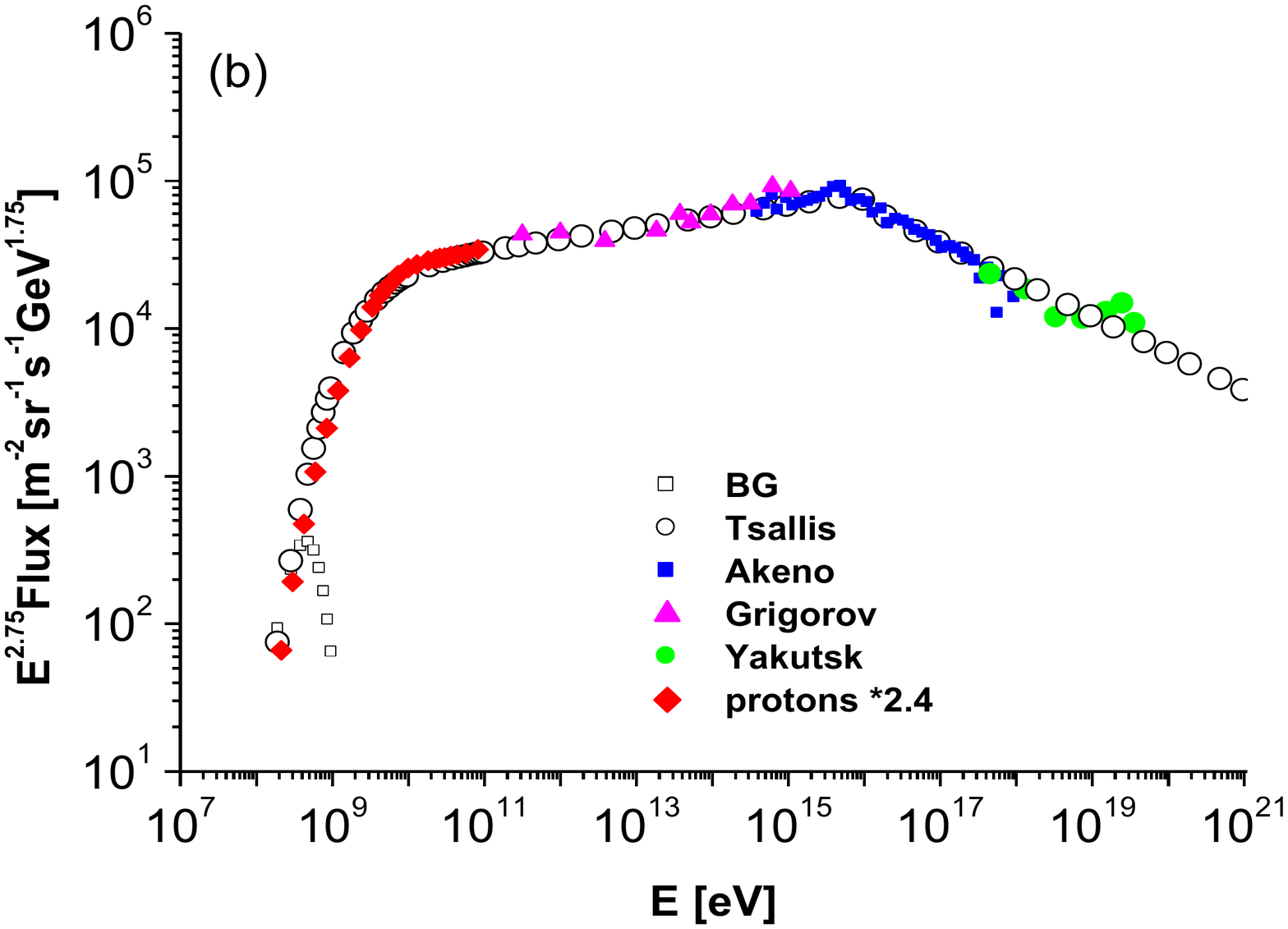}\vspace{-1.5cm}
\caption{(Color online) $(a)$ CR energy spectra fitted by the
single Tsallis distribution with $q=1.214$ and $T = 100$ MeV.
Notice the total inadequacy of a simple exponential (Boltzman,
denoted by BG here) distribution in description of these spectra.
$(b)$ The CR energy spectrum fitted by the double Tsallis
distributions discussed in text, cf. Eqs. (\ref{eq:fluxchange})
and (\ref{eq:alphai}). It was obtained for $T = 100$ MeV and
assuming that fluctuations of the temperature change abruptly at
$T_{cut} \approx 10^{15}$ eV $ \approx 10^{19}$ K from $q=1.214$
to $q=1.2$. Data are from \cite{SPS}, cf. also \cite{CR-origin}.}
\label{fig3}
\end{center}
\end{figure}

The above example tells us that it is reasonable to expect the
abrupt change of the specific heat in the CR source. Assuming that
this really happens and taking seriously the apparent connection
between $C_V$ and $q$ expressed by Eq. (\ref{eq:C_V}), we are lead
to the natural conjecture that in this case the usual fluctuation
pattern given by the gamma distribution (\ref{eq:Gamma}) should be
modifying accordingly. It is then assumed to be done by replacing
Eq. (\ref{eq:Gamma}) by its slightly modified version,
characterized by two nonextensivity parameters, $q_1$ (acting
before some temperature $T_{cut}$) and $q_2$ (acting after
$T_{cut}$). The change $q_1 \rightarrow q_2$ at $T_{cut}$ is
assumed to be abrupt and the temperature $T_{cut}$ becomes a new
parameter in our description. Following our proposition one
obtains the following flux of CR:
\begin{eqnarray}
\Phi(E) &=& N_0E^2\left[ \int_0^{1/T_{cut}} e^{-\beta
E}f_{q_1}(\beta)d \beta
+ \int_{1/T_{cut}}^{\infty} e^{-\beta E} f_{q_2}(\beta)d \beta \right] =\nonumber\\
&=& N_0\, E^2 \cdot \left[P_{q_1}(E) - \alpha_1 (E) P_{q_1}(E) +
\alpha_2(E) P_{q_2}(E)\right], \label{eq:fluxchange}
\end{eqnarray}
where $P_{q_i}(E)$  are given by Eq. (\ref{eq:P_q}) and
\begin{equation}
\alpha_i = \frac{\int_{1/T_{cut}}^{\infty} e^{-\beta E}
f_{q_i}(\beta)d\beta}{\int_0^{\infty} e^{-\beta E} f_{q_i}(\beta)
d\beta} = \frac{1}{\Gamma \left(\frac{1}{q_i -1}\right)}\Gamma
\left[\frac{1}{q_i-1}, \, \frac{1-\left(1-q_i\right)
E/T}{\left(q_i-1\right) T_{cut}/T}\right]. \label{eq:alphai}
\end{equation}
Our results are presented in Fig. \ref{fig3} where $q = 1.214$ in
Fig. \ref{fig3}a whereas in Fig. \ref{fig3}b $q_1=1.214$ and $q_2
= 1.2$; in both cases $T = 100$ MeV. Notice that now we do not
have a spectrum where, as in Section \ref{sec:IIB}, we change the
value of $q$ at some energy $E_c$ to get the observed structure.
Spectrum (\ref{eq:fluxchange}) is obtained by changing $q$ at some
temperature $T_{cut}$ (i.e., by changing slightly the shape of
gamma function (\ref{eq:Gamma})), this means that each $q$ gives a
spectrum for all energies. For this reason the parameters $q$ here
have slightly different values from these in Section
\ref{sec:IIB}. With spectrum given by Eq. (\ref{eq:fluxchange})
the "knee" region is reproduced very well, however, the price to
be paid is the need of a suitable choice of temperature at which
the fluctuation pattern changes (which amounts to assume the value
of $T_{cut} \approx 10^{15}$ eV $ \approx 10^{19}$ K)
\footnote{Two remarks: $(i)$ The nucleon superfluidity was
predicted already in \cite{Migdal} and today pulsar glitches
provide strong observational support for this hypothesis
\cite{NS}. Nucleon superfluidity arises from the formation of
Cooper pairs od fermions (actually in \cite{qSUP} also quark
superfluidity from cooling neutron stars were investigated).
Continuous formation and breaking of the Cooper pairs takes place
slightly below $T=T_C$ (critical temperature $T_C$ is in the order
$10^9 - 10^{10}$ K). $(ii)$ Neutron stars are born extremely hot
in supernova explosions, with interior temperatures around $T\sim
10^{12}$  K. Already within a day, the temperature in the cental
region of the neutron star will drop down to $\propto 10^9 -
10^{10}$ K and will reach $10^7$ K in about $100$ years
\cite{NeStar}. The first measurements of the temperature of a
neutron star interior (core temperature of the Vela pulsar is $T
\sim 10^8$ K, while the core temperature of PSR B0659+14 and
Geminga exceeds $2\cdot 10^8$ K) allow us to determine the
critical temperature $T_C \sim 7.5\cdot 10^9$ K \cite{critT}.}.

The above mechanism is only able to describe the "knee" region. To
describe all details seen in Fig. \ref{fig1}b let us consider an
other feature of heat capacity in Fermi liquids, namely its
dependence on the effective mass of nucleons consisting such
liquid. Following \cite{HC1} the proton heat capacity is
proportional to the ratio of the effective mass of the proton in
the neutron fluid to the mass of the free proton, $C\sim m^*/m$.
In the case of a mixture of Fermi liquids the proton effective
mass $m^*$ is affected by interactions with neutrons and other
protons and is given by
\begin{equation}
\frac{m^*}{m} = 1 + \frac{1}{3}D_p\left[ f^{pp}_1 + \left(
\frac{k_{F_n}}{k_{F_p}}\right)^2 f_1^{pn}\right],
\label{eq:mstaroverm}
\end{equation}
where $D_p$ denotes the density of quasiparticle states at the
Fermi surface given by wave vectors $k_{F_n}$ and $k_{F_p}$ for,
respectively, neutrons and protons, whereas $f^{pp}_1$ and
$f^{pn}_1$ are Landau parameters \cite{BJK}. Fig. \ref{fig1}b can
then be interpreted as showing changes of $C$ with energy in the
Fermi liquid. We start with the superfluid liquid with $C_1=4.7$
(here $m^*$ represents effective mass for $pp$ and $pn$
interactions),  when energy increases we stop to see nuclear
interactions and $C_2 = 5.1$ (with $m^*$ representing $pp$
interactions only), finally, for large $T$, one has the Fermi gas
with $C_3 = 5.3$ and, still further, the usual Fermi liquid
\footnote{It is worth to remember that fluctuations of temperature
we are talking about in this work refer to fluctuations in a small
region $V$. For a Fermi liquid the heat capacity expressed in
units of Boltzmann constant $k_B$ (i.e., for $k_B=1$) is of the
order $C\simeq 3\cdot 10^{35}$ cm$^{-3}$ \cite{YU}. Therefore,
taking values of $C$ estimated from the slope of the primary CR
spectra (cf. Fig. \ref{fig3}) one gets that the size of the region
of fluctuations is $V \sim 10^4$ fm$^3$.}. Notice that
\begin{equation}
\frac{1}{3}D_p f_1^{pp} = \frac{C_2-C_3}{C_3}~~~{\rm and}~~~
\frac{1}{3}D_p\left( \frac{k_{F_n}}{k_{F_p}}\right)^2f_1^{pn} =
\frac{C_1-C_2}{C_3}, \label{eq:estimations}
\end{equation}
this results in the following relation between Landau parameters,
\begin{equation}
\left( \frac{k_{F_n}}{k_{F_p}}\right)^2 \frac{f_1^{pn}}{f_1^{pp}}
= \frac{C_1 - C_2}{C_2 - C_3} = 2. \label{eq:LP}
\end{equation}
In the case of a one-component Fermi liquid we have the well known
identity, $m^*/m = 1 + F^{pp}_1/3$, where $F_1^{pp} =
D_pf_1^{pp}$. From (\ref{eq:LP}) we see that in a two-component
Fermi liquid the quantity $1-m^*/m$ is $3$ times bigger (this is
because the parameter $f_1$ which determines the interaction
between quasiparticles is negative, resulting in smaller effective
mass). From properties of excited states in nuclear matter ($Pb$
and neighboring  nuclei \cite{SZR}) $F_1^{np} = - 0.5 \pm 0.25$.
If $F_1^{nn} < F_1^{np} < F_1^{pp}$ and taking (after \cite{PH})
$F_1^{nn} - F_1^{pp} = -0.2$, we can estimate that for
neutron-star matter one has $m^*/m = 1 + F_1^{pp} \simeq 1 - 0.4
\pm 0.3 = 0.6 \pm 0.3$.

\subsection{\label{sec:IIIC}The notion of the effective
temperature $T_{eff}$}

Let us now come back to the results presented in Fig. \ref{fig3}.
Although a double Tsallis fit looks rather impressive there are
two shortcomings which we shall now discuss in more detail. First
is the fact that we still need a too high value of the
temperature, $T = 100$ MeV, which cannot be accommodated by any
reasonable physical mechanism of production of CR's. Second is the
very high value of the $T_{cut}$ temperature where change in the
fluctuation pattern is supposed to take place.

The possible way out of both dilemmas we are going to propose now
is to keep the value of $T = 100$ MeV but change its meaning. This
can be done by adding to the mechanism proposed in \cite{WW}
(which was accounting only for the possible fluctuations of $T$)
an additional effect of the possible viscosity which describes the
possible transfer of energy between the region of production and
surroundings (cf. \cite{WWTq} and \cite{hi} for details). As a
result one gets the same power-like distribution as before but
with the previous $T$ replaced by an {\it effective temperature} $
T_{eff}$ :
\begin{equation}
T_{eff} = T_0 + (q - 1) T_{visc}. \label{eq:Teff}
\end{equation}
Here  $T_0$ is the temperature around which one has fluctuations
and $T_{visc}$ is some new parameter depending on the transport
properties of the surrounding space around the emission region.

A few words of explanation are necessary at this point. Following
\cite{hi}, in the case of heavy ion collisions, where this concept
has been introduced for the first time, $T_{visc} = \eta
f(u)/(Dc_{\rho} \rho)$, where $\eta f(u)$ presents the effect of a
possible viscosity, with viscosity coefficient $\eta$ and $f(u) =
\left( \frac{\partial u_i}{\partial x_k} + \frac{\partial
u_k}{\partial x_l}\right)^2$ ($u$  being velocity), whereas $D$,
$c_{\rho}$ and $\rho$ are, respectively, the strength of the
temperature fluctuations, the specific heat under constant
pressure and the density. Here this quantity is supposed to model
the possible transfer of energy towards the CR particle. To
estimate its value let us notice that $(q - 1)T_{visc} \sim
(\tau/\tau_c)T_0$ where $\tau$ and $\tau_c$ are, respectively, a
relaxation time of the corresponding dissipative process and the
mean collision time. They have distinct physical meanings. Namely,
$\tau$ is the time (in most cases macroscopic) taken by the system
to spontaneously return to the steady state (whether in thermal
equilibrium or not) after being suddenly removed from it. It is,
to some extent connected to the mean collision time $\tau_c$ of
the particles responsible for the dissipative process. There is no
general formula linking them, their relationship depends in each
case on the system and circumstances under consideration. In
particular, $\tau$ is small for photon-electron and photon-photon
interactions at room temperature ($\propto 10^{-11}$ and $\propto
10^{-13}$ seconds, respectively). In the case of neutron star,
$\tau$ is of the order of the scattering time between electrons
(which carry the heat) but this fact is not obstacle (no matter
how large the mean free path of this electron may be) to consider
the neutron star as formed by a Fermi fluid of degenerate
neutrons. One finds in this case the relaxation times as large as
$\tau \sim 10^2$ seconds (for $T\sim 10^6$ K, $\rho \sim
10^{14}$g/cm$^3$ and $u \sim 10^3$ cm/s) \cite{3star} whereas in
the degenerate core of aged stars $\tau$ can reach $1$ second
\cite{1star}. The fact that $\tau$ can quantitatively greatly
differ from $\tau_c$ is most dramatically suggested by the
matter-radiation decoupling in the early universe (the relaxation
time of shear viscosity turns out to be several order of magnitude
larger than the collision time for most of the radiative era)
\cite{2star}. For this region one can expect that $T_{visc}$ may
be large enough (for example, $T_{visc}/T_0 \sim 10^8$) to be a
reasonably parameter describing the influence of surroundings on
the CR particle. Therefore in such picture there are fluctuations
of temperature $T$ characterized by the parameter $q$ and
connected with the character of the source, for example, as
discussed here, but there is also a surrounding space around  the
emission point which can pump energy into some selected region
from which a CR particle is emitted. This is described by the
positive parameter $T_{visc}$. As a result we are getting the
total $T_{eff}$ that grows with $q$.

Our schematic view in what concerns the fate of the CR spectrum is
then the following. $(i)$ CR are produced in an object where we
have low temperature which experiences some fluctuations around $T
= T_0$ given by $q > 1$ (with $T_0$ of the order of MeV). Now
$T_0$ can be as small as desired to properly fit the expected
features of the CR source. $(ii)$  By introducing to the previous
description \cite{WW} some (so far unspecified) flow
\cite{WWTq,hi}, the corresponding Tsallis distribution now has $T
= T_{eff}$, which can be quite large for example, of the order of
the observed $100$ MeV). $(iii)$ Therefore, in the spectrum of CR
one observes a Tsallis distribution with $T_{eff} \sim 100$ MeV
(as in the previous attempts \cite{previous1,previous2}) but now
it is not the temperature of the source of CR itself, but it is
composed from the temperature of the CR's source itself, $T_0$,
and the effect of the action of the surrounding space which is
given by $T_{visc}$. It should be stressed that such an effect
appears {\it only} in the case of Tsallis distribution. Only for
$q > 1$ is $T_{eff} > T_0$. If there are no temperature
fluctuations one always has $T_{eff} = T_0$.

Interestingly enough, the proposed mechanism seems to be capable
also to alleviate the second problem, namely the uncomfortably
high value of the $T_{cut}$ where the change of the nonextensivity
parameter $q$ occurs. This is because, as discussed in \cite{And},
when one considers viscosity effects in stars, one observes very
large abrupt changes of the viscosity coefficient which can easily
result in an effective $T_{visc}$ of the desired order to fit the
parameter $T_{cut}$.

\section{\label{sec:IV}Propagation and sources of cosmic rays}

Following \cite{previous1} one can argue that nonextensive
approach represented by Eq. (\ref{eq:P_q}) can also be connected
with the process of CR propagation. In our case such scenario
would mean that parameter $T_{visc}$ in $T_{eff}$ can represent
summarily some effective modelling of the propagation of CR.
Whereas detailed discussion of such a possibility is outside the
scope of the present paper, a few words on propagation of CR and
its possible influence on the spectra and composition of CR are in
order here.

So far we have considered only the sources of CR insisting on the
possibility that they are of thermal but nonextensive type. We
shall now address the possible influence on our results by the
propagation of CR. Measurements of the composition and energy
spectra of CR characterize the CR population after the observed
particles have travelled from their production sites (i.e., the
{\it sources}) through distant space towards the remote detectors.
It is expected that both the composition of the particles and the
shape of their energy spectra undergo in such process changes due
to a variety of processes encountered during the propagation.

The results presented in this work concern only production in the
source, which is supposed to be thermal and nonextensive for some
reasons discussed here. However, the fact that one can confront
them with experimental data indicates the possibility that they
depend only weakly on the propagation process in what concerns the
slopes of the energetic spectrum and the global characteristics of
the chemical composition (notwithstanding the fact that
propagation is very important factor).

The possible way out of this dilemma is to argue that the break in
the original spectrum and connected with it phase transition
occurs actually at much slower energies and that resultant
spectrum is then accelerated to the observed energies - for
example by the magneto-hydrodynamical turbulence and/or shock
discontinuities (i.e., by the so called diffusive shock
acceleration (DSA) mechanism, cf. \cite{DSA}). The simplest
version of this mechanism, as discussed in \cite{FM}, implies that
distribution function after the shock, $f_{after}(p)$, is related
to the original distribution before the shock, $f_{before}(p)$ in
the following way:
\begin{equation}
f_{after}(p) = \frac{b}{p^b}\, \int_{p_{min}}^p dp'\, p'^{(b-1)}\,
f_{before}(p'), \label{eq:DSA}
\end{equation}
where $b=3r/(r-1)$. Here $p$ denotes the particle momentum,
$r=\rho_2/\rho_1$ describes the compression of densities across
the shock and $p_{min}$ denotes the minimal value of momenta. DSA
mechanism transforms a $\delta\left(p-p_0\right)$ spectrum of
relativistic particles in a power-like spectrum of the type $f(p)
\propto p^{-b}$. For example, if one has initial spectrum of the
form $\propto p^{-c}$ which encounters a shock with strength given
by $b$ then Eq. (\ref{eq:DSA}) shows that: $(i)$ in the case of $b
> c$ (which corresponds to the initial spectrum injected into the
shock being softer than it would result from a $\delta$-function)
one has $f_{after} \propto p^{-c}$ (i.e., the acceleration process
does not change the shape of the spectrum); $(ii)$ in the case of
$b<c$ (i.e., for the steep initial spectrum) one has $f_{after}
\propto p^{-b}$, which coincides with result of injection of a
$\delta$-function into the shock. It can be then shown from Eq.
(\ref{eq:DSA}) that when the strength of shocks is larger then the
slope of the injected spectrum, the shape of the spectrum should
be given by the production spectrum in source \cite{FM}.

To summarize this part: stochastic mechanisms of acceleration of
CR particles (like acceleration on the fronts of shock waves or
Fermi acceleration in turbulent plasmas, both analogous in some
sense to Brownian motion) do not change the shape of the
power-like production spectra, but, unfortunately, they are not
particularly effective, i.e., they do not lead to large increase
of energy \cite{Gaisser}. The increase of energy per one collision
is of the order $\Delta E/E \sim u^2/c^2$, what for the plasma
velocity $u \approx 10^6 - 10^7$ cm/s gives $\Delta E/E \approx
10^{-8}$ and leads to the mean relative increase of energy during
the time life of Galaxy ($t\approx 3\cdot 10^{17}$ s) only by
factor $\langle \delta\rangle \approx 3$. Fluctuation on the steep
spectrum of accelerated particles result in additional increase of
energy. Because of the multiplicative character of acceleration we
have log-normal distribution of variable $\delta$, $P(\delta) d\ln
\delta = \left(\sqrt{2\pi}\sigma \right)^{-1} \exp\left[-\left(
\ln \delta - \ln \langle \delta \rangle \right)^2/2\sigma^2
\right] d\ln \delta$, what results in shift of the spectrum of
source on the energy scale by $\langle \delta \rangle ^{(\gamma
-1)/\gamma} \exp\left[(\gamma - 1)^2\sigma^2/(2\gamma)\right]$,
where $\sigma^2$ is variation of the distribution $P(\delta)$. For
$\gamma \approx \langle \delta \rangle \approx \sigma^2 \approx 3$
we can obtain only order of magnitude shift of the energy
spectrum.

Diffusive propagation of CR component is commonly summarized in a
continuity equation for the differential density, $N_i(E)$, of
each component \cite{Diff}. If convection effects are neglected
(which is probably valid approximation at high energies) together
with effects due to energy gain or loss and to radioactive decay,
the continuity equation becomes \cite{Cont}
\begin{equation}
N_i(E) = \frac{1}{\Lambda^{-1}(E) + \Lambda^{-1}_s(A)}\left[
\frac{Q_i(E)}{\beta c \rho} + \Sigma_{k>1}\frac{N_k}{\Lambda_{k
\to i}}\right]. \label{eq:diff}
\end{equation}
Here $Q_i(E)$ is the rate of production in the source and
$\Lambda_{k \to i}$ quantifies the probability of a nucleus $k$ to
spallate into a product $i$ in an interstellar interaction. The
two quantities, the propagation length, $\Lambda (E) = \beta c
\rho \tau(E) $, and the average spallation length, $\Lambda_s(A)=
m/\sigma(A)$, characterize the propagation of cosmic rays and the
change of the atomic number (the loos of nuclei) due to the
spallation effect (here $\beta$ is velocity, $\rho$ the density of
the material in the galactic space and  $\tau(E)$ is the average
time which particle spend in the galaxy).

The propagation path length decreases with energy but it is
assumed that it has the same value for different nuclei of the
same rigidity. On the other hand, the spallation path length
depends on the atomic number $A$ (essentially like $\Lambda_s
\propto A^{-2/3}$), its energy dependence remains, however, very
weak for the relativistic particles and is therefore neglected.
For high energies (above TeV/amu) $\Lambda < \Lambda_s$ and for
the approximate scaling behavior $\Lambda_{k \to i} = \Lambda_{k
\to i}\left( A_i/A_k \right)$ the relative abundance is roughly
given by
\begin{equation}
\frac{N\left(A_i,E\right)}{N\left( A_k,E\right)} \propto
\frac{Q\left(A_i,E\right)}{Q\left(A_k,E\right)}.
\label{eq:abundance1}
\end{equation}
For stable nuclei the observed abundances do not differ
substantially from the relative abundances in the source
\cite{Cont}.

Some remarks concerning the applicability of Eq. (\ref{eq:diff})
are in order here. It is frequently assumed that the propagation
path length $\Lambda$ decreases as function of energy, $\Lambda
\propto E^{-0.6}$ \cite{propath}. Since the interaction length is
almost independent of the primary energy this necessitates $Q(E)
\propto E^{-2.1}$ spectrum at the sources to explain the observed
$N(E) \propto E^{-2.7}$ spectrum at the Earth. Because of this,
the values of $q$ evaluated by us from fits to the observed
spectrum  must be regarded as the corresponding lower limits of
the nonextensivity parameter $q$ ($q \approx 1.213$ for $\gamma =
2.7$ and $q \approx 1.244$ for $\gamma = 2.1$). In fact, because
of large uncertainty in what concerns the energy dependence of the
propagation path length which still exists, we cannot present
exact values of $q$ in the source. For example, the model
presented in \cite{Ber} predicts flatter than the above mentioned
spectra at the sources before the {\it knee} requiring therefore a
stronger dependence of $\Lambda (E)$ on energy. Recent
measurements of the TeV gamma ray flux from a shell type supernova
remnant yield spectral index $\gamma = 2.19\pm 0.09\pm 0.15$
\cite{Ah}, in agreement with the standard model \footnote{The
Standard Model for Galactic Cosmic Rays is based on Supernova
Remnant (SNR) paradigm and includes four basic elements: $(i)$
SNRs as the sources, $(ii)$ SNR shock acceleration, $(iii)$
rigidity dependent injection as mechanism providing the observed
CR mass composition and $(iv)$ diffusive propagation of CR in the
galactic magnetic fields \cite{BV}.}. On the other hand, for the
Crab Nebula a steeper spectrum (with $\gamma = 2.57\pm 0.05$) has
been obtained \cite{CM}, indicating that probably not all sources
exhibit the same behavior. Moreover, the $\Lambda \propto
E^{-0.6}$ dependence of the propagation path length cannot be
extrapolated to the {\it knee} energies whereas taking $\Lambda
\propto E^{-0.2}$ dependence, as discussed in \cite{propath},
necessitates additional assumptions concerning the spectral shape
at the source, $Q(E)$, in order to explain the observed spectra
with spectral indices in the range $\gamma \sim 2.55 \pm 2.75$.

\section{\label{sec:V}Summary and conclusions}

We have proposed and discussed the possibility that CR can
originate from nonextensive thermal sources described by a
nonextensive formalism proposed in \cite{Tsallis}. Our motivation
was the observation that the spectrum of CR has, in general, a
power-like shape, $E^{-\gamma}$, and such behavior is naturally
accounted for in a nonextensive approach with nonextensivity
parameter $q=(3+\gamma)/(2+\gamma)$. Looking more closely one
encounters a characteristic "knee" structure in this power-like
behavior, with $\gamma_1 \simeq 2.7$ at energies below $E_{knee}
\sim 10^{15}$ eV and $\gamma_2 \simeq 3.1$ above it. This can be
also accounted for in a nonextensive approach with two different
values of nonextensivity parameter: $q_1 = 1.213$ before the
"knee" and $q_2 = 1.196$ above it.

From our previous experience with applications of the
nonextensivity to different physical processes (cf., \cite{WWTq}
and references therein) we can trace the origin of such power-like
behavior back to some intrinsic, nonstatistical fluctuations of
temperature in the CR's source \cite{WW,WWTq,Add}. In this case
the nonextensivity parameter $q$ is regarded as a measure of the
heat capacity (see Eq. (\ref{eq:C_V})). This means that the
measured energy spectrum (Fig. \ref{fig1}a) can be converted to
the energy dependence of the heat capacity $C$. The result is
shown in Fig. \ref{fig1}b. As one can see, $C_V$ acts here as a
kind of magnifying glass converting all subtle structures of
$\Phi(E)$ into much more pronounced and structured bumps. Its
importance would parallel the long-standing discussion of the
origin of the "knee"-like structure of the energy spectrum, but
exposed in a much more dramatic and visible way.

As a plausible physical mechanism leading to changes in $C_V$ of
the order of $C_2/C_1 = 1.09$ (corresponding to change in $q$,
describing the observed change in the spectral index $\gamma$) we
have proposed fluid/superfluid transitions in Fermi liquids used
to describe neutron stars, which we model by suitably modifying
the gamma distribution (\ref{eq:Gamma}) describing temperature
fluctuations (essentially by changing parameter $q$ at some
temperature $T_{cut}$).  To get fits as presented in Fig.
\ref{fig3}, while at the same time keeping temperature in the CR's
source, $T_0$, acceptably low (of the order of MeV, i.e. of the
order of the interior stars temperature) we have to resort to an
approach allowing not only for fluctuations of temperature but
also for the energy transfer to the production region from its
surroundings introduced recently in \cite{WWTq,hi}. This allows us
to keep the critical temperature (corresponding to the nucleon
superfluidity) around $T_C \sim 0.1 - 1$ MeV with effective
temperature used in the fits remaining as high as $T_{eff} = 100$
MeV. This also allows to quantitatively understanding that the
origin of changes of the nonextensivity parameters at the
temperature as high as $T_{cut} \simeq 10^{15}$ eV $\simeq
10^{19}$ K required in our approach could be in some specific
viscous effects in stars \cite{And}.

One should stress at this point that the mechanism we proposed,
namely that CR can indeed originate from nonextensive thermal
sources, must be, for a while, regarded only as a plausible
scheme, which would have to be checked together with other
mechanisms aiming to describe the CR spectra, their composition
and propagation \cite{CR-origin}. The need for such an analysis
(which, however, goes outside the limited scope of this paper) is,
for example, visible when one realizes the following. On one hand
one observes that the energy spectrum of CR depends mainly on the
nonextensivity parameter $q$. Dependence on $T$ is visible only
for low energies. For $E >> T$ we observe scale-free, mostly
$T$-independent behavior. The temperature has therefore marginal
influence on the shape of the energy spectrum of CR. On the other
hand, a quite opposite situation is encountered when considering
the chemical composition of CR where $\langle A\rangle$ depends
only very weakly on $q$ but it depends linearly on $T = T_{eff}$,
see Eq. (\ref{eq:Comp}). It is then plausible that analyzing
simultaneously the energy spectrum and composition one could
obtain both $q$ and $T_{eff}$ (i.e., according to Eq.
(\ref{eq:Teff}), the $T_0$ of the source and $T_{visc}$
responsible for the energy transfer). Notice that around the
"knee" (where one expects changes in the chemical composition) one
can essentially freely vary $T_{eff}$ without affecting the shape
of the energy spectrum but substantially changing the chemical
composition $\langle A\rangle$.

\section*{Acknowledgements}

Partial support (GW) of the Ministry of Science and Higher
Education under contract 1P03B02230 is gratefully acknowledged.

\end{document}